\documentclass{jfm}
\usepackage{amssymb,amsmath,amsfonts,graphicx}

\newcommand{\eq}[1]{\begin{equation} #1 \end{equation}}

\newcommand{\const}{\mathop{\rm const}\nolimits}
\newcommand{\p}{\partial}

\title[Acceleration of hydrodynamic vortices]{Acceleration of hydrodynamic vortices in open systems}

\author[E. A. Pashitskii, V. N. Malnev, and R. A. Naryshkin]%
{E.\ns A.\ns P\ls A\ls S\ls H\ls I\ls T\ls S\ls K\ls I\ls I$^1$,%
  V.\ns N.\ns M\ls A\ls L\ls N\ls E\ls V$^2$\break
\and R.\ns A.\ns N\ls A\ls R\ls Y\ls S\ls H\ls K\ls I\ls N$^{2,3}$ \thanks{E-mail: rnaryshk@uwo.ca },\ns}

\affiliation{$^1$Institute of Physics of NAS of Ukraine,  46 Prosp. Nauki, 03028 Kiev, Ukraine\\[\affilskip]
$^2$Taras Shevchenko Kiev National University, 6 Prosp. Glushkova,
03022 Kiev, Ukraine\\[\affilskip]
$^3$Department of Applied Mathematics, University of Western Ontario, \\London, Ontario N6A 5B7, Canada}

\pubyear{1993} \volume{538} \pagerange{1--10}
\date{26 Feb 2007}
\begin{document}

\maketitle

\begin{abstract}
A new class of exact solutions of hydrodynamic equations for an
incompressible fluid (gas) at the presence of a bulk sink and
uprising vertical flows of matter is considered.
The acceleration of the rotation velocity of classical
non-stationary vortices is conditioned by the joint action of the
convective and Coriolis hydrodynamic forces (accelerations), which
appear due to the converging radial flows of the matter in the
region of a bulk sink. It is
shown that there exist velocity profiles that nullify viscous terms in the Navier-Stokes
equations and represent a vortex structure with a "rigid-body"\ rotation of its core and converging
radial flows.

The concept of non-stationary vortices in open systems is applied
to description of origination of power atmospheric vortices
(whirlwinds, tornados, and typhoons). In the classical hydrodynamics
a favorable condition for the origination and existence of such
vortices is the exact nullification of the terms, which describe kinematic
viscosity of an incompressible fluid. Such flows have the minimal rate of energy
dissipation that corresponds to the "minimum entropy
production principle", and therefore may relatively easily appear in
favorable natural conditions.
\end{abstract}

\section{Introduction}

Hydrodynamical vortices and vortex-like flows are extremely
widespread in nature. This is evidently proven by constant
occurrence of powerful atmospheric vortices --- cyclones,
hurricanes, typhoons, tornados (see \cite{1,2}), sandstorms, oceanic
vortical flows --- the so-called "rings", twirled streams
(whirlpools) on rivers, turbulent vortices in wake swirls of
ships, etc. An example of a huge stationary vortical structure is the
big "Red Spot"\ on the surface of Jupiter, which has been observed by
astronomers for several hundred of years.

One may ask a natural question: why such hydrodynamic vortex
structures can exist for a long time in liquids and gases despite the
finite viscosity of the medium? The answer to this question is partly
given by the Rankine model (see \cite{3}) for a vortex in an
incompressible viscous fluid where its azimuthal velocity profile is set
to be linear in the radius $r$ of a "rigid-body"\ rotation $v_\varphi (r)
= \omega r $ with the angular velocity $ \omega $ inside a certain
cylindrical region $r\leqslant R_0 $ of a radius $R_0,$ and
have the "differential"\ rotation $v_\varphi (r) = \omega R^2_0/r $ in the
external region $r> R_0.$ In this case, the terms with a bulk viscosity
in the Navier-Stokes equations for an incompressible liquid (see
\cite{4}) are precisely nullified in the cylindrical system of
coordinates. At the first sight, this corresponds to the non-dissipative
vortical rotation of a liquid with a constant value of
$\omega=\mathrm {const} $ and with a sharp break of a profile
$v_\varphi (r) $ at the point $r=R_0. $

However, the radial dependence of the velocity $v_\varphi (r) $ of real
hydrodynamical vortices cannot have any singular points (in
particular, jumps or breaks that correspond to jumps of the first
derivative) and also must be a continuous analytical function of the
radius $r.$ Therefore, in a neighborhood of the point $r=R_0,$ the
function $v_\varphi (r) $ should be smooth, and  there is a
finite viscosity in this area, which results in dissipation of kinetic
energy of a vortex, its spreading in space and decaying in time.
Stationary or accelerating vortices can exist only due to inflow of
energy from the external environment. At this, from all possible
hydrodynamical flows and vortical movements in an incompressible
liquid, easily arise, longer exist and dominate over
others only those ones, whose viscosity effects are almost absent and energy dissipation
plays an insignificant role. Such hydrodynamical flows and vortices
correspond to the "minimum of entropy
production  principle"\ and are frequently enough observed in the nature.

In the present paper we will consider several examples of
origination of non-stationary vortical structures in open
multicomponent systems (solutions, homogeneous and heterogeneous
mixtures), in which due to chemical or phase
transformations a bulk sink (convergence) of the matter exists
when one or several components drop out from the general collective
hydrodynamical motion of the system. In this case, it is supposed that
the chemical compound of the substance, as well as its density $ \rho =\mathrm {const}, $
are constant in time and
almost homogeneous in space due to dynamical and chemical
balance with the surrounding environment.

We will show, that the growth of the rotation velocity of classical
non-stationary vortices is caused by the joint action of convective
and Coriolis hydrodynamic forces (accelerations) that arise due
to converging radial flows of the substance flowing to the area of a bulk
sink. The law of the velocity growth in time can be exponential,
or can correspond to a nonlinear "explosive"\ regime of instability,
when  velocities formally reach infinite values in a finite time
interval (under the condition of unlimited inflow of the
substance from the environment).

A limit of the growth of velocity and kinetic energy of a vortex
in a normal liquid (gas) in real conditions is caused by several
factors: the friction between the liquid (or gas) and fixed solid surfaces;
the development of turbulent dissipative processes, in particular, on the
border of the vortex core at $r=R_0, $ where a tangential jump of the
velocity arises, and a connected with it small-scale instability of
surface waves; the decrease of inflow of a substance from the outside,
etc. We shall also note that at approaching of hydrodynamical
velocity to the speed of sound,
the compressibility of a liquid must be taken into account,
which in turn accounts for dissipation
due to the volumetric viscosity. All these factors will be considered
and estimated below, but before that a simple mechanism of origination
of a vortex in an incompressible one-component liquid (gas) due to
existence of parallel to the vortex axis hydrodynamic flows, whose
velocity depends on the longitudinal coordinate $z,$ will be
considered. Such a mechanism may explain the formation of a funnel in
water or a sand-spout in a desert due to accelerated falling of
the water in the hole or rising of the hot air in the gravitation
field of the Earth.

\section{One-component liquid with accelerated flows:
the mechanism of formation of a funnel and a windspout}

One of the most interesting paradoxes in hydrodynamics is the
so-called "funnel effect"\ (see \cite{3,8}). It is usually assumed, that
this effect is caused by conservation laws of the angular
momentum of an incompressible fluid (gas) inside a given contour,
and is accompanied by the accelerated rotation of a vortex at
concentration of the vorticity of a flow $ {\omega} = \mathrm {rot} \,
\mathbf {v} $ due to the narrowing of the channel. Another approach
to the problem of a funnel formation lies in the assumption (see
\cite{9}) about origination of the angular momentum at zero
initial vorticity as a result of instability of
cylindrically-symmetric flow in a liquid (the flooded jet) in
relation to axially-asymmetric left- and right-spiral perturbations
with ejection of rotation of a certain sign to the infinity at
the expense of a flow (convective instability) and accumulation of a
rotation motion of the other sign (absolute instability).

Let us show, that there exists one additional simple mechanism of a vortex
formation in an incompressible liquid (gas), which is present in a
gravitational field and includes vertical ascending or descending
flows, whose velocities depend on the coordinate $z$ (along the
vertical axis of the vortex).

Consider the Navier-Stokes equations for an axially-symmetric motion
of an incompressible viscous fluid (gas) in cylindrical
coordinates (see \cite{4}):
\begin{equation} \label{1.1}
\begin{array}{l}
\displaystyle \frac{\partial v_r} {\partial t} + v_r\frac{\partial
v_r} {\partial r }-\frac{v_\varphi^2} {r} = \displaystyle-\frac{1}
{\rho} \frac{\partial P} {\partial r} + \nu\left(\frac{\partial^2
v_r} {\partial r^2} + \frac{1} {r} \frac{\partial v_r} {\partial r}-
\frac{v_r} {r^2} \right),
\end{array}
\end{equation}
\begin{equation} \label{1.2}
\begin{array}{l}
\displaystyle \frac{\partial v_\varphi} {\partial t} +
v_r\frac{\partial v_\varphi} {\partial r}+\frac{v_r v_\varphi } {r}
= \nu\left(\frac{\partial^2 v_\varphi} {\partial r^2} + \frac{1} {r}
\frac{\partial v_\varphi} {\partial r}- \frac{v_\varphi} {r^2}
\right),
\end{array}
\end{equation}
\begin{equation} \label{1.3}
\begin{array}{l}
\displaystyle \frac{\partial v_z} {\partial t} + v_r\frac{\partial
v_z} {\partial r} +v_z\frac{\partial v_z} {\partial z} =
\displaystyle-\frac{1} {\rho} \frac{\partial P} {\partial z}
 - g +\nu\left(\frac{\partial^2 v_z} {\partial
r^2} + \frac{1} {r} \frac{\partial v_z} {\partial r} +
\frac{\partial^2 v_z} {\partial z^2} \right),
\end{array}
\end{equation}
where $v_r, v_\varphi $ and $v_z $ are radial, azimuthal and axial
components of the hydrodynamical velocity $ \mathbf {v}, $ $P $ and $
\rho $ are the pressure and the density of a liquid (gas), $\nu =\eta/\rho $
is the coefficient of kinematic viscosity, and $g $ is the acceleration due to gravity,
which is directed to the opposite direction of the axis $z. $

Equations (\ref{1.1})--(\ref{1.3}) are completed with the
continuity equation
\begin{equation}
\label{1.4} {\mathrm{div} \, \mathbf{v}} = \frac{\partial v_r}
{\partial r} + \frac{v_r} {r} + \frac{\partial v_z} {\partial z} =0.
\end{equation}
Let us notice, that in equations (\ref{1.1}) and (\ref{1.2}),
dependencies of $v_r $ and $v_\varphi $ on $z$ are not taken into
account only for simplification.

In the case of a plane vortical rotation, when $v_r=v_z=0, $ equations
(\ref{1.1}) and (\ref{1.2}) become:
\eq{\frac{v^2_\varphi} {r} = \frac{1} {\rho} \frac{d P} {d r},
\label{1.1a}}
\eq{\frac{\partial v_\varphi} {\partial t} = \nu\left(
\frac{\partial^2 v_\varphi} {\partial r^2} + \frac{1} {r}
\frac{\partial v_\varphi} {\partial r}- \frac{v_\varphi} {r^2}
\right) .\label{1.2a}}

If the radial dependence of the azimuthal velocity $v_\varphi (r), $
according to the Rankine  model of a vortex (see \cite{3}), is chosen to be
\eq{v_\varphi (r) = \left\{\begin{array}{c} \omega r, \; \; \qquad
r\leqslant R_0, \\ \omega R_0^2/r, \quad r> R_0, \end{array}
\right.\label{1.5}}
then the right hand side of equation (\ref{1.2a}) is identically equal
to zero, which corresponds to the rotation of an incompressible
liquid (gas) with a constant angular velocity $ \omega =\mathrm
{const}. $

At this, the distribution of hydrodynamical pressure $P (r), $
according to (\ref{1.1}), is in the so-called `cyclostrophic' rotation
regime and has the form:
\eq{P (r) = \left\{\begin{array}{c} P_0 +\rho\omega^2r^2/2, \; \;
\qquad r\leqslant R_0, \\ P_\infty-\rho\omega^2R_0^4/2r^2, \quad r>
R_0, \end{array} \right.\label{1.6}}
where $P_0 = P_\infty-\rho\omega^2R_0^2 $ is the pressure on the
vortex axis, and $P_\infty $ is the pressure at large distances (i.e. at
$r \to \infty $).

In the Rankine model, the radius of the vortex core $R_0 $ is not
determined, but if in a liquid (gas) there  exists a cylindrically-symmetric flow
(the flooded jet) with a velocity $v_z, $ then its radius
will determine the size $R_0 $ in equation (\ref{1.5}). Let us assume, that
the flow velocity depends linearly on the coordinate $z$ and does
not depend on $r $ in the region $r\leqslant R_0, $ i.e. we have the profile:
\eq{v_z (z) = \left\{\begin{array}{c} v_{z0} + \alpha z, \quad
r\leqslant R_0, \\ 0, \qquad\qquad \; r> R_0.\end{array} \right.
\label{1.7}}
In this case the continuity equation (\ref{1.4}) is satisfied for
the following radial dependence of velocity $v_r, $ which is continuous at
$r=R_0: $
\eq{v_r (r) = \left\{\begin{array}{c}\displaystyle-\frac12 \alpha r,
\, \quad r\leqslant R_0, \\\displaystyle-\frac12 \alpha \frac{R^2_0}
{r}, \quad r> R_0.\end{array} \right.\label{1.8}}
Let us notice, that the above mentioned structure of the velocities
(\ref{1.5}), (\ref{1.7}) and (\ref{1.8}) identically nullify the viscous terms in
equations (\ref{1.1})--(\ref{1.3}). At the same time, the diagonal
components of the viscous stress tensor  are
different from zero for these profiles, which results in the following expression for the
rate of changing of the kinetic energy of a vortex due to dissipation (per unit of
length of a vortex along the axis $z $):
\eq{ \label{1.11a} \left( \frac{dE_{\mathrm{kin}}}{dt}
\right)_{\mathrm{dis}} =-4\pi\rho\nu \left[ \frac32\alpha^2
+\omega^2(0) \right] R_0^2. }

Substituting expressions (\ref{1.5}) and (\ref{1.8}) in equation
(\ref{1.2}), we get
\eq{\frac{d\omega} {dt} = \left\{\begin{array}{c} \alpha\omega, \,
\quad r\leqslant R_0, \\ 0, \qquad r> R_0.\end{array} \right.
\label{1.9}}
Equation (\ref{1.9}) is a result of that fact that in the inner region
($r\leqslant R_0$) the convertive and Coriolis forces at $v_r\ne0$
are added up, while in the external region ($r>R_0$) they mutually
compensate each other.

If the parameter $ \alpha $ is constant in time $ (\alpha =\mathrm
{const}, \;\displaystyle \frac{d\alpha} {dt} =0) $ and positive $
\alpha> 0, $ then from equation (\ref{1.9}) it follows that inside
the region $r\leqslant R_0 $ the angular velocity of a vortex grows in
time according to the exponential law
\eq{\omega (t) = \omega (0) \, e^{\alpha t}, \label{1.12}}
provided that a nonzero initial vorticity $ \omega (0) \ne 0 $ exists
in the liquid (gas), whereas in the external region $ \omega =\omega (0)
= \const. $ Thus, on the border of a vortex core $r=R_0, $ a jump of the
azimuthal velocity that exponentially grows in time arises (see
section \ref{turb}).

Let us emphasize, that the energy dissipation (\ref{1.11a}) of the
non-stationary vortical motion does not depend on time $t,$ and is
determined only by the initial vorticity $2\omega (0). $ Thus, the
dissipation remains small, despite the fast increase of the angular
velocity of a "rigid-body"\ rotation of the vortex core. This means
that non-stationary vortices are not suppressed by the dissipation
at least at an initial stage of their developments.

On the other hand, substituting expressions (\ref{1.5}), (\ref{1.7}) and
(\ref{1.8}) in equations (\ref{1.1}) and (\ref{1.3}), we get the
following equations for the determination of the hydrodynamical pressure:
\eq{\frac{\partial P} {\partial r} = \left\{\begin{array}{c}
\displaystyle\rho r\left[\omega^2 (t)-\frac{\alpha^2 (t)} {4} +
\frac12\frac{d\alpha} {dt} \right], \, \qquad\;\;\; r\leqslant R_0,
\\\displaystyle\frac {\rho R^4_0} {r^3} \left[\omega^2 (0) + \frac{\alpha^2 (t)}
{4} + \frac12\frac{d\alpha} {dt} \frac{r^2} {R^2_0} \right], \quad
r> R_0,
\end{array} \right.\label{1.10}}
\eq{\frac{\partial P} {\partial z} = \left\{\begin{array}{c}
\displaystyle-\rho\left[g +\alpha v_{z0} +z\left(\frac{d\alpha} {dt}
+ \alpha^2\right) \right], \; \quad r\leqslant R_0,
\\\displaystyle-\rho g, \qquad\quad\qquad\qquad\qquad\qquad\qquad r>
R_0.\end{array} \right.\label{1.11}}

Let us notice, that equations (\ref{1.10}) and (\ref{1.11}) indicate
on the existence and increasing in time of a jump of the first
derivatives of the pressure $ \p P/\p r $ and $ \p P/\p z $ on the
surface of the vortex core $r=R_0. $ From equation (\ref{1.11}), a
possibility of existence of a non-stationary solution with $
\frac{d\alpha} {dt} \ne0$ follows, namely:
\eq{\frac{d\alpha} {dt} + \alpha^2 (t) =0, \quad \frac{\partial P}
{\partial z} \pm \rho g=0.\label{1.13}}
The first equation (\ref{1.13}) at $ \alpha> 0 $ has the following solution:
\eq{\alpha (t) = \frac{\alpha_0} {1 +\alpha_0 t}, \quad
\alpha_0\equiv \alpha (0)> 0, \label{1.14}}
whereas the second equation of (\ref{1.13}) corresponds to the hydrostatic
pressure distribution in the whole space, and the sign (+)
corresponds to an ascending flow (the axis $z $ is directed
upwards), while the sign ($-$) corresponds to a descending flow (the axis $z $
is directed downwards). In this case, equation (\ref{1.9}) in the
region $r\leqslant R_0 $ has a solution
\eq{\omega (t) = \omega (0) \exp\left\{\int_0^t\frac{\alpha_0 \, dt
'} {1 +\alpha_0 t '} \right\} =\omega (0) (1 +\alpha_0 t),
\label{1.15}}
which corresponds to the linear growth of vortex rotation velocity  in time.

At last, we notice that equation (\ref{1.9}) under the condition
of $ \alpha (t) = \omega (t)> 0 $ takes the form:
\eq{\frac{d\omega} {dt}-\omega^2 (t) =0. \label{1.18}}
The solution of the nonlinear equation (\ref{1.18}) corresponds to
the so-called "explosive"\ instability:
\eq{\omega (t) \equiv\alpha (t) = \frac{\omega (0)} {1-\omega (0)
t}, \label{1.19}}
when in a finite time interval $t_0=1/\omega (0) \equiv 1/\alpha (0)
$ the angular velocity of a fluid rotation $ \omega (t) $ and the
derivative of the axial velocity with respect to $z, $ i.e. $ \alpha (t) \equiv \p
v_z/\p z, $ formally approach infinity, although they are actually
limited from the above due to the effects of compressibility of a liquid.

The positive sign of $ \alpha $ corresponds to the growth of
velocity of the flow along the axis $z$ directed by velocity $v_{z0}.$
Thus, in a descending flow of a liquid which flows out through a
hole at the bottom under the action of gravitation and accelerates
along the axis $z$ by the linear law (\ref{1.7}), we get an
exponential (\ref{1.12}), linear (\ref{1.15}), or "explosive"\
(\ref{1.19}) laws of the acceleration of vortex rotation in time. At
this, the amount of a liquid, which flows out, is completely
compensated by the inflow of the same amount of the liquid with
velocity of the converging radial flow (\ref{1.8}) from the
surrounding region, which is considered as a large enough reservoir
of the substance.

Such a simple model explains the funnel formation in a bath at the
opening of a hole or whirlpool formation on a river in the
place of a sharp deepening of the bottom. The deceleration of the rate of
rotation velocity growth of a liquid (water) in the
stationary regime is caused by the friction with fixed solid
surfaces as well as by the energy dissipation on the tangential jump
of the azimuthal velocity $v_\varphi $ at the point $r=R_0 $ (see
section \ref{turb}).

This model may also explain the origination of sandy tornados in
deserts. Due to a strong heating  of some sites of a
surface of a sandy ground by the sunlight (the darkest or located perpendicularly to
the solar rays), the nearby air gets warm locally and starts rising
upwards with acceleration under the action of the Archimedean force
as less heavy. If this acceleration in quasi-stationary conditions
 corresponds approximately to the linear in $z$ law (\ref{1.7}), then at
$ \alpha> 0 $ we again obtain the exponential law of rotation
velocity growth of a vortex (\ref{1.12}). At this, the accelerated
decrease in time of the pressure near the vortex axis $r=0$ with the increase of
$\omega (t) $ and $ \alpha (t) $ leads to the suction of
the sand deep into the vortex and to formation of visible tornados, which
are frequently observed in deserts.


\section{Vortices in multi-component open systems with chemical and phase transformations}

Consider a multi-component system in which chemical processes or
phase transitions are taking place. Partial balance equations of
substances for each component (phase) look like (see \cite{8}):
\eq{\frac{\partial \rho_i} {\partial t} + {\mathrm{div}} \,
(\rho_i{\mathbf {v}}_i) =Q_i, \quad \frac{dM_i} {dt} = \int_V Q_i \,
dV, \label{2.1}}
where $ \rho_i $ and $ \mathbf {v}_i $ are density and
hydrodynamical velocity of the $i$-th component, $M_i $ is the mass of
this component in a volume $V, $ $Q_i $ is the capacity of a source
($Q_i> 0 $) or sink ($Q_i <0 $) of this component due to chemical reactions
or phase transformations.

For a closed system the conservation law of the total mass of all
$N$ components holds true:
\eq{M =\sum_{i=1}^NM_i ={\mathrm{const}}, \quad \sum_{i=1}^NQ_i=0,
\label{2.2}}
and therefore the general continuity equation can be written as
\eq{\frac{\partial \rho}{\partial
t}+{\mathrm{div}}(\rho{\mathbf{v}})=0, \quad
\rho=\sum_{i=1}^N\rho_i,\quad
\rho{\mathbf{v}}=\sum_{i=1}^N\rho_i{\mathbf{v}}_i.\label{2.3}}

For open systems with a bulk sink or source of the substance and its
unlimited inflow from the environment under conditions of dynamic and
chemical balance, one or several components  can appear in the system or drop out
from the general collective motion of the matter (liquid or gas) as a result of chemical
reactions or phase transformations.
At the same time, diffusion and hydrodynamical flows support constant density and
chemical potential in the open system, i.e.
under stationary conditions it can be with a good approximation assumed
that $ \rho =\mathrm {const}. $ In this case the effective
continuity equation can be written in the form
\eq{{\mathrm{div} }\, {\mathbf {v} }=Q/\rho\ne 0, \label{2.4}}
where $Q $ is a certain capacity of a source ($Q> 0 $) or sink ($Q
<0 $) of the substance in a system or in a some part of its volume
$V. $

Further we shall consider systems with a homogeneous in space bulk
sink of matter in a some finite cylindrical region with the
radius $R_0. $ In cylindrical coordinates, equation (\ref{2.4}) looks
like (at the absence of longitudinal flows, $v_z=0 $):
\eq{{\mathrm {div}} \, {\mathbf{v}} \equiv \frac{\partial v_r}
{\partial r} + \frac{v_r} {r} = \left\{\begin{array}{c} - |Q
|/\rho\equiv-1/\tau , \quad r\leqslant R_0, \\ 0, \qquad\qquad\qquad
\; \; \; \; r> R_0.\end{array} \right.\label{2.5}}
From the last expression it follows that the radial velocity can be set as
\eq{v_r (r) = \left\{\begin{array}{c}-\beta r, \qquad\quad
r\leqslant R_0, \\-\beta R^2_0/r, \quad\; r> R_0, \end{array}
\right.\label{2.6}}
where $ \beta=1/2\tau> 0. $ We notice, that $ |Q |, \;\tau $ and $
\beta $ can be either constants, or variables in time, but must be
homogeneous in space, i.e. the condition $ \nabla \mathrm {div} \,
\mathbf {v} =0,$ which is necessary and sufficient for the use of the
Navier-Stokes equations (\ref{1.1}) and (\ref{1.2}) for an
incompressible liquid, is satisfied.

Substituting expressions (\ref{1.5}) and (\ref{2.6}) for azimuthal
and radial velocity profiles in equations (\ref{1.1}) and (\ref{1.2}), we
get the equation for determination of hydrodynamic pressure (compare with
(\ref{1.10})):
\eq{\frac{\partial P} {\partial r} = \left\{\begin{array}{c}
\displaystyle\rho r \left(\omega^2-\beta^2 +\frac{d\beta} {dt}
\right), \; \; \qquad r\leqslant R_0, \\\displaystyle\frac {\rho
R^4_0} {r^3} \left(\omega^2 +\beta^2 +\frac{d\beta} {dt} \frac{r^2}
{R^2_0} \right), \quad r> R_0,
\end{array} \right.\label{2.7}}
and also the equation for the angular velocity of rotation of the
liquid (gas) in a vortex (compare with (\ref{1.9})):
\eq{\frac{d\omega} {dt} = \left\{\begin{array}{c} 2 \beta \omega,
\quad r\leqslant R_0, \\ 0, \; \qquad r> R_0.\end{array}
\right.\label{2.8}}

In the case of a constant sink $ |Q | =\mathrm {const} $ and $ \beta
=\mathrm {const} $ we get from (\ref{2.8}) the exponential law of
acceleration of a vortical rotation (at $ \omega (0) \ne 0 $):
\eq{\omega (t) = \omega (0) \, e^{t/\tau} .\label{2.9}}

If the capacity of the surrounding reservoir is finite, and the
velocity of inflow (and also the sink) of a substance in the dynamic
equilibrium state decreases in time, for instance, by the exponential
law
\eq{\beta (t) \equiv \frac12\frac{|Q (t) |} {\rho} \equiv\beta_0 \,
e^{-t/t_0}, \quad \mathrm{i.e.} \quad \tau (t) = \tau_0 \,
e^{t/t_0}, \label{2.10}}
then the time dependence of the angular velocity of a vortex
rotation, according to (\ref{2.8}), looks like
\eq{\omega (t) = \omega (0) \, \exp\left\{\frac{t_0} {\tau_0}
\left(1-e^{-t/t_0} \right) \right\}\label{2.11}}
and tends to the maximum value $ \omega_\infty =\omega (0) \, e
^{t_0/\tau_0} $ at $t\to\infty, $ whereas during the time $t\ll t_0
$ the exponential growth of velocity $ \omega (t) \approx \omega (0)
\, e^{t/\tau_0}$ remains.

From equation (\ref{2.8}) it follows, that at $ \beta> 0 $ and either
sign of $ \omega,$ there exists another non-stationary solution at $
\beta =\frac12 | \omega (t) |, $ when equation (\ref{2.8}) becomes
(for $r\leqslant R_0 $):
\eq{\frac{d |\omega |} {dt}-\omega^2 (t) =0.\label{2.12}}
The solution to this equation corresponds to the nonlinear
instability of the "explosive"\ type (compare with (\ref{1.18})):
\eq{| \omega (t) | = \frac{| \omega (0) |} {1-|\omega (0) |t}
.\label{2.13}}
For a finite interval of time $t_0=1 / |\omega (0) |,$  the angular
velocity $ \omega (t) $ as well as $ \beta (t) $ formally reach their
infinite values.

Thus, the presence of a bulk sink together with the unlimited inflow of a
substance in open systems with chemical reactions or phase transformations, due
to nonlinear hydrodynamical forces, which arise under the action of
converging flows $v_r <0, $ results in the acceleration of vortices
even at the absence of axial flows ($v_z=0 $). An example of such a natural
mechanism of vortex acceleration due to chemical reactions can be origination of
the so-called "fiery tornados", which arise during big fires
in closed volumes at the formation, on the one hand, of solid
combustion materials, in particular solid compounds of oxygen
(oxides) that completely exclude oxygen from the air, and, on the
other hand, at the presence of free inflow (draught), which creates
radial and ascending flows of the external air enriched with oxygen.

Moreover, during the dissolution of solid crystals in a liquid (for
example, manganic-sour potassium in water), when the heavier
solution is immersed downwards by gravity, and a new pure solvent with
origination of converging radial flows goes to its place, a
hydrodynamical vortex may arise, which due to friction will force
the dissolving crystals to rotate.

\section{Origination of tornados and typhoons during formation of dense cloud systems}
\label{sec4}

Let us consider one more concrete example of origination of vortices
in an open non-equilibrium heterogeneous system with phase
transformations, namely, in a humid atmosphere during water
vapor condensation in rain clouds where tornados and typhoons might
originate.

Despite the colossal quantity of available observable data on
tornados and typhoons in natural conditions and numerous attempts of
modeling these powerful atmospheric vortices in laboratory
plants and computer simulations (see \cite{2}), the true reasons for
origination and development of such phenomena are not yet understood
completely. The majority of theoretical models of tornados
and typhoons can be reduced to the well-known
"funnel effect"\ in hydrodynamics (see \cite{3,8}).
At this, it is supposed that the
radial squeeze of twirled descending flows of dense humid air is
being realized by the air masses, which are flowing in the region of low
atmospheric pressure.

In Ref. (\cite{5}) an essentially new mechanism of tornado and typhoon
origination during formation of dense cloud systems was introduced.
This the mechanism is directly connected to the process of intensive
condensation of water vapor at cooling of the humid air below the
dew-point. A "vapor -- liquid"\ phase transition with the formation of
weighed water droplets in a cloud (fog) is accompanied by a large
decrease (about 800 times) of the specific volume which is
occupied by water molecules. This means that the condensation of vapor
should result in essential (almost twice at initial $100 \% $
humidity of the air) decrease of the density of a gaseous component of the
two-phase heterogeneous system "air -- water drops"\ in comparison
with the initial density of a chemically homogeneous gas mixture "air --
water vapor". It is obvious that this is equivalent to the existence of
a bulk sink (convergence) of a substance inside the cloud and
origination of the concentration gradient of water molecules on the
border of a cloud.

As it follows from the Boltzmann equation for a mixture of gases with
close by weight molecules (see \cite{10}), the diffusion flow of one
of the components under the action of the corresponding concentration
gradient, as a result of intermolecular collisions, creates an
average hydrodynamical flow of the gas mixture as a whole. Since the cloud is a
thermodynamically non-equilibrium open system with the unlimited
inflow of a substance from the outside, the presence of a bulk
sink and the humidity gradient caused by the condensation of moisture
should result in appearance of converging hydrodynamical flows of
the humid air from the surrounding areas of the atmosphere. Under
conditions of a dynamical balance, such flows retain a constant in time
and almost homogeneous in space average density $ \rho =\mathrm
{const},$ which is equivalent to the condition of incompressibility of
the medium.

As has been shown in the previous section, within the framework
of the hydrodynamical model of an incompressible viscous liquid with a
bulk sink (convergence) of the substance and converging radial flows,
under the joint action of convective and Coriolis forces an instability of
a vortical "rigid-body"\ rotation appears, which can evolve in time
either by the exponential law (in the case of constant sink and
inflow of substance), or by a scenario of the "explosive"\ type when
the infinite rotation velocity is reached in a finite time interval
(under the condition of simultaneous unlimited increase of
velocities of the sink and the inflow).

We will show, that such an
instability can be one of the reasons of origination and acceleration
of powerful tornadoes and typhoons during the formation of dense
cloud systems (see also \cite{5}).

\subsection{Balance of substance during the condensation of rain clouds}

The rain cloud represents an open non-equilibrium two-phase system
(a mixture of the air and a fog) that self-organizes due to condensation
of water vapor and keeps small water drops in a certain volume
by converging flows of the air and the Stokes friction forces.

Consider a cloud of a cylindrical form of the radius $R_0$ and height $h. $ As
inside a forming cloud the process of vapor condensation from the humid
air, cooled below the dew-point $T_{\mathrm {dew}}$ (at the given
pressure distribution $P$) takes place with the certain velocity,
the effective balance equation of the substance inside a cloud, with account of
horizontal (radial) and vertical (ascending or descending) air
flows from the surrounding atmosphere, and a bulk sink of the
substance with capacity $Q <0,$  can be written as
\eq{2\pi R_0h v_r (R_0) + \pi R^2_0 [v_z (h)-v_z (0)] =-\frac{|Q |}
{\rho} \pi R^2_0 h.\label{3.1}}
Substituting expressions (\ref{1.7}) and (\ref{1.8}) in equation
(\ref{3.1}), we get the relation
\eq{2\beta =\alpha +\frac{|Q |} {\rho}, \label{3.2}}
which corresponds to the effective continuity equation (compare with
(\ref{1.4}) and (\ref{2.5}))
\eq{\frac{\partial v_r} {\partial r} + \frac{v_r} {r} +
\frac{\partial v_z} {\partial z} =-\frac{|Q |} {\rho}
\equiv-\frac{1} {\tau}.\label{3.3}}

Thus, at $ \alpha> 0 $ there is a strengthening of the nonlinear
instability, which results in acceleration of a vortex rotation in
the region $r\leqslant R_0 $ by the following law:
\eq{\omega (t) = \omega (0) \exp\left\{\int_0^tdt '\left[\alpha (t
') + \frac{1} {\tau (t ')} \right] \right\}. \label{3.4}}
At the same time, in the external region $r> R_0, $ in agreement with
(\ref{2.8}), the angular velocity remains constant and is equal to
the initial value $ \omega (0). $ This means that on the border of the
vortex core at $r=R_0 $ a tangential discontinuity of azimuthal velocity
arises and grows in time. This jump at $ \alpha =\mathrm {const} $
and $ |Q | =\mathrm {const} $ equals
\eq{\Delta v_\varphi (R_0, t) \equiv V_0 (t) = \omega (0) R_0 [e
^{2\beta t}-1], \label{3.5}}
and in the case of an "explosive"\ instability we have
\eq{V_0 (t) = \omega (0) R_0\left[\frac{1} {1-\omega (0) t}-1\right]
.\label{3.5b}}

\subsection{Instability of the tangential jump of velocity on the border of the tornado core}
\label{turb}

As is well-known (see \cite{4}), a tangential jump of velocity in
hydrodynamics results in origination of instability of surface
perturbations of an incompressible fluid. The dispersion equation
of small perturbations in linear approximation with account of a
finite viscosity looks like (see \cite{5}):
\eq{(\omega-kV_0) (\omega-kV_0+i\nu k^2) =-\omega (\omega+i\nu k^2),
\label{3.6}}
where $k=2\pi/\lambda $ is the wave number of surface waves with the
length $\lambda,$ which are distributed along a flat surface. The
equation (\ref{3.6}) with a good approximation can be also applied
to the case of the cylindrical surface of a radius $R_0 $ if we consider
$kR_0\gg 1. $

The frequency $ \omega_0 (k) \equiv \mathrm {Re} \; \omega (k) $ and
the increment $ \gamma (k) \equiv \mathrm {Im} \; \omega
(k) $ of such short-wave surface perturbations on the tangential
jump of velocities with transversal displacements of a surface along
the $z$ axis  and the propagation along the $x $ axis  (at $k> 0 $)
\eq{\zeta (x, z, t) = \zeta_0 \, e^{-k|z |} \, e^{\gamma t} \, \exp
(ikx-i\omega_0t), \label{3.7}}
according to dispersion equation (\ref{3.6}), are equal to
\eq{\omega_0 (k) = \frac{kV_0} {2}, \quad\gamma (k) =
\frac12\left[\sqrt {k^2V_0^2 +\nu^2k^4}-\nu k^2\right] .\label{3.8}}

If $k\ll V_0/\nu $ (but $kR_0\gg1 $), then the increment is $ \gamma
(k) =kV_0/2. $ At $k\gg V_0/\nu, $ according to (\ref{3.8}), the
increment tends to its maximum value (at $k\to\infty $):
\eq{\gamma_\infty=V_0^2/2\nu.\label{3.9}}

It is worth noting, that relations (\ref{3.6}) -- (\ref{3.9})
were obtained under the assumption of a constant value of the velocity
jump $V_0 =\mathrm {const}. $ But if the condition
$\gamma_\infty\gg2\beta$ is satisfied, i.e. if the jump of the velocity
(\ref{3.5})  grows in time very slowly (adiabatically) in comparison
with development times of the unstable short-wave excitations with
increment (\ref{3.9}), it is possible to  present approximately the
time evolution of the amplitudes of surface perturbations at $z=0$
in the form
\eq{| \zeta (t) | \simeq\zeta_0 \, e^{\gamma_\infty t} = \zeta_0
\,\exp\left\{\frac{\omega^2 (0) R_0^2t} {2\nu} \left(e^{2\beta
t}-1\right)^2\right\}\label{3.10}}
for the exponential instability of a vortex (for $ \beta =\const $),
or
\eq{| \zeta (t) | = \zeta_0 \,\exp\left\{\frac{\omega^2 (0) R_0^2t}
{2\nu} \left[\frac{1} {1-\omega (0)
t}-1\right]^2\right\}\label{3.10b}}
in the case of the "explosive"\ instability given by (\ref{1.18}) and
(\ref{2.13}).

During the time interval $t\ll1/2\beta,$ when $V_0 (t) \simeq
2\beta\omega (0) R_0t, $  we have
\eq{| \zeta (t) | \simeq\zeta_0 \,\exp\left\{\frac{2\beta^2\omega^2
(0) R_0^2t^3} {\nu} \right\},\label{3.11}}
instead of the "exponent to the exponent"\ law (\ref{3.10}).

A more detailed analysis of time evolution of small surface
perturbations at a variable velocity $V_0 (t) $ can be obtained in
the case where $ \nu=0 $ with the help of the following linear
differential equation:
\eq{\frac{d^2\zeta} {dt^2}-kV_0 (t) \frac{d\zeta} {dt} +
\frac12k^2V_0^2 (t) \zeta (t) =0.\label{3.12}}
If $V_0 (t) \sim t, $ then the dominate time dependence of the solution
(\ref{3.12}) looks like
\eq{\zeta (t) \sim\exp\left\{at^2-bt\right\},\label{3.13}}
i.e. it is slower, than the dependence (\ref{3.11}). But if one
takes into the account the fact, that at $ \nu\to0 $ the inequality $k\ll V_0
(t)/\nu$ always takes place, i.e. the increment is $ \gamma=kV_0
(t)/2, $ then instead of (\ref{3.11}) we get (at $2\beta t\ll 1 $):
\eq{| \zeta (t) | \simeq\zeta_0 \,\exp\left\{\beta\omega (0)
kR_0t^2\right\},\label{3.14}}
i.e. the use of the "adiabatic expressions"\ (\ref{3.10}) is possible at early stages
of a vortex instability evolution. Similarly, in the
case of "explosive"\ instability, with the account of (\ref{3.5b})
and (\ref{3.10b}) under the condition of $ \omega (0) t\ll 1,$ we obtain
\eq{| \zeta (t) | \simeq\zeta_0 \,\exp\left\{k\omega^2 (0)
R_0t^2\right\}.\label{3.14b}}

\subsection{Development of the turbulence on the border of the tornado core}

As is known, the maximum value of the growth increment of unstable perturbations in
a some non-homogeneous layer of thickness $l$ corresponds to the
wave number $k_m\simeq 1/l.$ In the case of surface waves on a
tangential jump of velocity, the role of the thickness of a transition
non-homogeneous layer is played by the doubled amplitude of these waves
(surface displacement), i.e. $k_m ^{(t)} \simeq 1/2 |\zeta (t) |. $
Substituting this value in expression (\ref{3.14}), we get the
equation
\eq{\ln\frac{| \zeta (t) |} {\zeta_0} \simeq \frac{\beta \omega (0)
R_0 t^2} {2 |\zeta (t) |}, \label{3.15}}
from which it follows that $ | \zeta (t) | $  changes approximately
according to the law $ | \zeta (t) | \sim t^2/\ln t. $ At this, the effective
coefficient of the anomalous turbulent viscosity inside a surface
layer of thickness $2 |\zeta (t) | $ can be estimated as
\eq{\nu^* (t) \simeq \zeta^2 (t) \gamma (t) \simeq\frac12\beta\omega
(0) R_0 |\zeta (t) | t, \label{3.16}}
i.e. the value $ \nu^* (t) $ grows in time, according to
(\ref{3.15}), almost by the cubic law ($ \sim t^3 $) and can reach
very large values ($ \nu^*\gg\nu $).

The saturation of turbulent perturbations occurs due to the nonlinear
dissipation, when the increase of the kinetic energy of surface waves,
as a result of instability of the tangential jump of velocity
$\frac12\gamma\rho \tilde v^2, $ is compensated by the turbulent
dissipation energy per unit time and unit volume
$\rho\tilde v^3/l $ (see \cite{4}), where $ \tilde v\simeq d |\zeta
(t) |/dt $ is the velocity of turbulent pulsations, $l\simeq |\zeta
(t) | $ is a characteristic scale of turbulence, and $
\gamma\simeq\pi V_0 / |\zeta | $ is the maximum increment of the
instability. Therefore, it follows that
\eq{\tilde v (t) \simeq\frac{\pi} {4} V_0 (t).\label{3.17}}

On the other hand, the saturation of acceleration of the angular
velocity of a "rigid-body"\ rotation of the vortex core (tornado)
occurs when the azimuthal velocity $v_\varphi (t) =R_0 \omega (t)$
and the value of a velocity jump $V_0 (t)$ reach velocities close to the
speed of sound in the air $c_s\simeq 330 $ m/s, i.e. when the
compressibility effects of the air and the effects of a finite bulk
viscosity appear. The characteristic time of  a rotation saturation of the
vortex, according to (\ref{3.5}), equals
\eq{t_{\mathrm {max}} \simeq\frac{1} {2\beta} \ln\left[\frac{c_s}
{\omega (0) R_0} \right] .\label{3.18}}

Thus, $ \tilde v_{\mathrm {max}} \simeq c_s $ and $ \gamma_{\mathrm
{max}} \simeq\pi c_s/\zeta_{\mathrm {max}}, $ where the value of
$\zeta_{\mathrm {max}}, $ according to (\ref{3.15}), is determined
by the equation:
\eq{\ln\frac{\zeta_{\mathrm{max}}} {\zeta_0} \simeq
\frac{t^2_{\mathrm{max}} \beta\omega (0) R_0}
{2\zeta_{\mathrm{max}}}. \label{3.19}}
Let us notice that in the case of the "explosive"\ regime of acceleration
 of a vortex, it is necessary to replace $\beta $ by $ \omega (0) $
 in expressions (\ref{3.15}), (\ref{3.16}) and (\ref{3.19}),
and the value $t_{\mathrm {max}} $ under the condition $c_s\gg \omega
(0) R_0 $ equals $t_{\mathrm {max}} \simeq 1/\omega (0). $

If we assume that the initial azimuthal and radial velocities of air on
the borders of a cloud, which has radius $R_0\simeq 1 $ km, are equal by
order to $v_0 =\omega (0) R_0 =\beta R_0\simeq 10 $ m/s, i.e. $
\beta \simeq 10^{-2} $ s$^{-1},$ then the time of acceleration of the
tornado core to the speed of sound $c_s $ equals $t_{\mathrm
{max}} \simeq (1\div2) \cdot 10^2 $ s for the "explosive"\ or
exponential acceleration laws of a tornado, respectively. Setting the minimum
initial value of displacement $ \zeta_0 $ of a surface $r=R_0$ to
the free path of molecules in the air at the normal pressure $l_0\simeq 5\cdot
10^{-7} $ cm, we get from (\ref{3.19}) the following
estimation for the maximum scale of the turbulent pulsations:
$\zeta_{\mathrm {max}} \simeq (20\div 80) $ m.

In this case the maximum value of the turbulent viscosity in the
surface layer of a thickness $2\zeta_{\mathrm {max}}$ on the border
of a vortex core $r=R_0 $ equals
\eq{\nu^*_{\mathrm {max}} \simeq\frac{1} {2} \omega (0)
R_0\zeta_{\mathrm {max}} t_{\mathrm {max}} \beta\label{3.20}}
and can by many orders exceed the usual coefficient of the kinematic viscosity
of the air. In particular, for the above mentioned estimations
$t_{\mathrm {max}} $ and $ \zeta_{\mathrm {max}} $ we get the value
$\nu^*_{\mathrm {max}} \simeq 5\cdot 10^6 $ cm$^2$/s, whereas the
usual kinematic viscosity of the air equals $ \nu\simeq 0.15 $
cm$^2$/s, i.e. their ratio is $ \nu^*_{\mathrm {max}}/\nu \simeq
3\cdot 10^7. $

Thus, on the surface of the central part of a powerful tornado of a
radius $R_0,$ which rotates almost with  velocity of sound
$V_{\varphi {\mathrm {max}}} = \omega (0) R_0 \,\exp
[\gamma_{\mathrm {max}} t_{\mathrm {max}}] \simeq c_s, $ a very
viscous layer of a thickness $2\zeta_{\mathrm {max}} \ll R_0 $ and
with the viscosity $ \nu^*_{\mathrm {max}} \gg \nu$ appears because of  the
turbulence. Due to this fact, a partial entrainment of the nearby air in
the region $r> R_0 $ occurs, and therefore the profile of azimuthal
velocity distribution of the air in this region changes. We will seek
it in the form
\eq{v_\varphi (r, t) = \Omega (r, t) {R_0^2} / {r} \quad (r> R),
\label{3.21}}
where $ \Omega (r, t) $ is an unknown function of $r $ and $t, $
which, according to (\ref{1.2}), satisfies the equation:
\eq{\frac{\partial \Omega} {\partial t}-\frac{\beta R_0^2} {r}
\frac{\partial\Omega} {\partial r} = \nu^*_{\mathrm {max}}
\left(\frac{\partial^2\Omega} {\partial
r^2}-\frac1r\frac{\partial\Omega} {\partial r} \right) \label{3.22}}
with a boundary condition $ \Omega (R_0, t) = \omega (t) = \omega
(0) \, e^{2\beta t}. $ From here we get the following expression for the
rotation velocity of the air in the external region $r> R_0: $
\eq{v_\varphi (r, t) = \omega (0) \, R_0 \, e^{2\beta t}
\left(\frac{R_0} {r} \right)^{\sigma+1} \frac{K_\sigma\left({r} /
{l^*}\right)} {K_\sigma\left({R_0} / {l^*}\right)}, \label{3.23}}
where
$K_\sigma (x) $ is the McDonald function (the modified Bessel function), $ \sigma = ({R_0} /
{2l^*})^2-1, $ and $l^* =\sqrt {\nu^*_{\mathrm {max}}/2\beta}. $ Using the obtained above
value of $ \nu^*_{\mathrm {max}},$ we get the following
estimation: $l^*\simeq 150 $ m.

On large distances $r\gg l^* $ the velocity (\ref{3.23}) changes
according to the law
\eq{v_\varphi (r) \sim \left(\frac{R_0} {r} \right)^{\sigma+1} \sqrt
{\frac{l^*} {r}} \; e^{-r/l^*},\label{3.24}}
that explains rather a weak motion of the air around a tornado on the
distances of $r> 150 $ m. On the other hand, those fact that the turbulent
pulsations and rotation velocity of a vortex core can reach the
velocity of sound ($ \tilde v_{\mathrm {max}} \simeq
V_{\varphi\mathrm {max}} \simeq c_s $) explains the reason of the
intensive generation of low-frequency sound waves in powerful
tornadoes (a "roar"\ of a tornado).

\subsection{Formation of a tornado funnel with account of gravity
and vertical flows of the air}

As was mentioned above, for the flows of the air, which flow into a rain cloud
of a cylindrical form during the time of its condensation, the following
hydrodynamical velocities are natural:
\begin{equation}
\begin{array}{l}
\displaystyle v_r =
\left\{\begin{array}{l}-\beta r, \! \! \! \! \qquad\quad r\leqslant R_0, \\
-\beta R_0^2/r, \! \! \! \! \; \quad r> R_0, \\ \end{array}
\right.\quad v_\varphi =
\left\{\begin{array}{l} \omega r, \! \! \! \! \qquad\quad r\leqslant R_0, \\
\omega R_0^2/r, \! \! \! \! \; \quad r> R_0, \\ \end{array} \right.
\quad v_z =
\left\{\begin{array} {rcl} v_{z0} + \alpha z, \! \! \quad r\leqslant R_0, \\
0, \! \! \quad r> R_0, \\ \end{array} \right. \label{3.25}
\end{array}
\end{equation}
where the parameter $ \beta $ is determined by relation
(\ref{3.2}). In the case of exponential instability we have $ \omega
(t) = \omega (0) \, e^{2\beta t} $ (for $ \alpha =\mathrm {const} $
and $ |Q | =\mathrm {const} $). At this, bulk viscous forces in
equations (\ref{1.1})--(\ref{1.3}) are equal identically to zero, so
that we get
\begin{equation}
\displaystyle \frac{\partial P} {\partial r } =
\left\{\begin{array}{l} \displaystyle\rho r\left[
\omega^2 (t)-\beta^2\right], \; \; \, \quad r\leqslant R_0, \\
\displaystyle\frac{\rho R_0^4} {r^3} \left[\omega^2 (0) +
\beta^2\right], \quad r> R_0, \\ \end{array} \right. \label{3.26}
\end{equation}
\begin{equation}
\frac{\partial P} {\partial z} =
\left\{\begin{array}{l}\displaystyle-\rho\tilde
g-\rho\alpha^2z, \; \; \tilde g=g +\alpha v_{z0}, \; \; r\leqslant R_0, \\
\displaystyle-\rho g, \qquad\qquad\qquad\qquad\qquad \; \; r>
R_0.\end{array} \right. \label{3.27}
\end{equation}

Integration of equations (\ref{3.26}) and (\ref{3.27})
determines the difference of the pressure between two arbitrary points:
\begin{equation}
\begin{array}{l}
\displaystyle P_2-P_1 =\displaystyle \left\{\begin{array}{l}
\displaystyle-\rho\tilde g (z_2-z_1)- \frac{\rho} {2} \; \alpha^2
(z_2^2-z_1^2) +  \frac{\rho} {2} \left[\omega^2 (t)-\beta^2\right]
(r_2^2-r_1^2), \; \, \qquad
  r_{1,2} \leqslant R_0, \\
\displaystyle-\rho g (z_2-z_1) - \frac{\rho R_0^4} {2} \left[
\omega^2 (0) + \beta^2\right] \left(\frac{1} {r_2^2}-
\frac{1} {r_1^2} \right),\qquad\qquad\quad \qquad r_{1,2}> R_0. \\
\end{array} \right. \label{3.28}
\end{array}
\end{equation}
From (\ref{3.28}) it follows, that the form of a surface of
constant pressure (an isobar), which corresponds to the point of
water drops evaporation $P _{\mathrm {evp}}$ in the internal region
$r\leqslant R_0 $ is determined by the equation
\begin{equation}
\begin{array}{l}
\displaystyle z^2 (r, t) + \frac{2\tilde g} {\alpha^2} z (r, t) +
\frac{R^2_0} {\alpha^2} \left[ \omega^2 (t) + \omega^2 (0) \right]
\displaystyle-\frac{r^2} {\alpha^2}
 \left[\omega^2 (t)-\beta^2\right] - \frac{2\left(
P_\infty-P_{\mathrm {evp}} \right)} {\rho\alpha^2} =0, \label{3.29}
\end{array}
\end{equation}
where the coordinate $z $ is counted from the initial flat surface
$P_0=P_\infty=P_{\mathrm {evp}}, $ and in the external region $r> R_0$
is determined by the relation
\begin{equation}
\begin{array}{l}
\displaystyle z (r, t) =z_0 (t)-\frac{\left[\omega^2 (0) +
\beta^2\right] R_0^4} {2 g r^2} + \frac{\left( P_\infty-P_{\mathrm
{evp}} \right)} {\rho g}, \label{3.30}
\end{array}
\end{equation}
where the function $z_0 (t) $ is to be obtained from the condition of the
isobar continuity at the point $r=R_0. $ According to equation
(\ref{3.29}), we find the value of the coordinate $z (r, t) $ at the point
$r=R_0: $
\begin{equation}
\begin{array}{l}
\displaystyle z (R_0, t) =-\frac{\tilde g} {\alpha^2}\displaystyle
+\sqrt {\frac{\tilde g^2} {\alpha^4}-\frac{\left[ \omega^2 (0) +
\beta^2\right] R_0^2} {\alpha^2} + \frac{2\left( P_\infty-P_{\mathrm
{evp}} \right)} {\rho\alpha^2}} = \const, \label{3.31}
\end{array}
\end{equation}
so that
\begin{equation}
\begin{array}{l}
\displaystyle z_0=z (R_0) + \frac{R_0^2} {2 g} \left[ \omega^2 (0) +
\beta^2\right]-\frac{\left( P_\infty-P_{\mathrm {evp}} \right)}
{\rho g} = \const, \label{3.32}
\end{array}
\end{equation}
i.e. in the external region $r> R_0,$ the form of the isobar does not
depend on time.

The coordinate of a point of the isobar on the vortex axis $r=0, $
according to (\ref{3.29})--(\ref{3.32}), is determined by the
expression
\begin{equation}
\begin{array}{l}
\displaystyle z (0, t) =-\frac{\tilde g} {\alpha^2} \displaystyle +
\sqrt {\frac{\tilde g^2} {\alpha^4}-\frac{R_0^2\left[ \omega^2 (t) +
\omega^2 (0) \right]} {\alpha^2} + \frac{2\left( P_\infty-P_{\mathrm
{evp}} \right)} {\rho\alpha^2}}. \label{3.33}
\end{array}
\end{equation}

From the last formula it follows, that for $ \tilde g> 0,$ with the growth of the
angular velocity $ \omega (t) $ as a result of the exponential
instability, there is an increase of the absolute value of the
negative coordinate $z (0, t), $ which corresponds to the deepening
of the minimum of the function $z (r, t) $ in the region $z <0. $

\begin{figure}
\begin{center}
  \includegraphics[height=0.6\textheight,width=\textwidth,angle=-90]{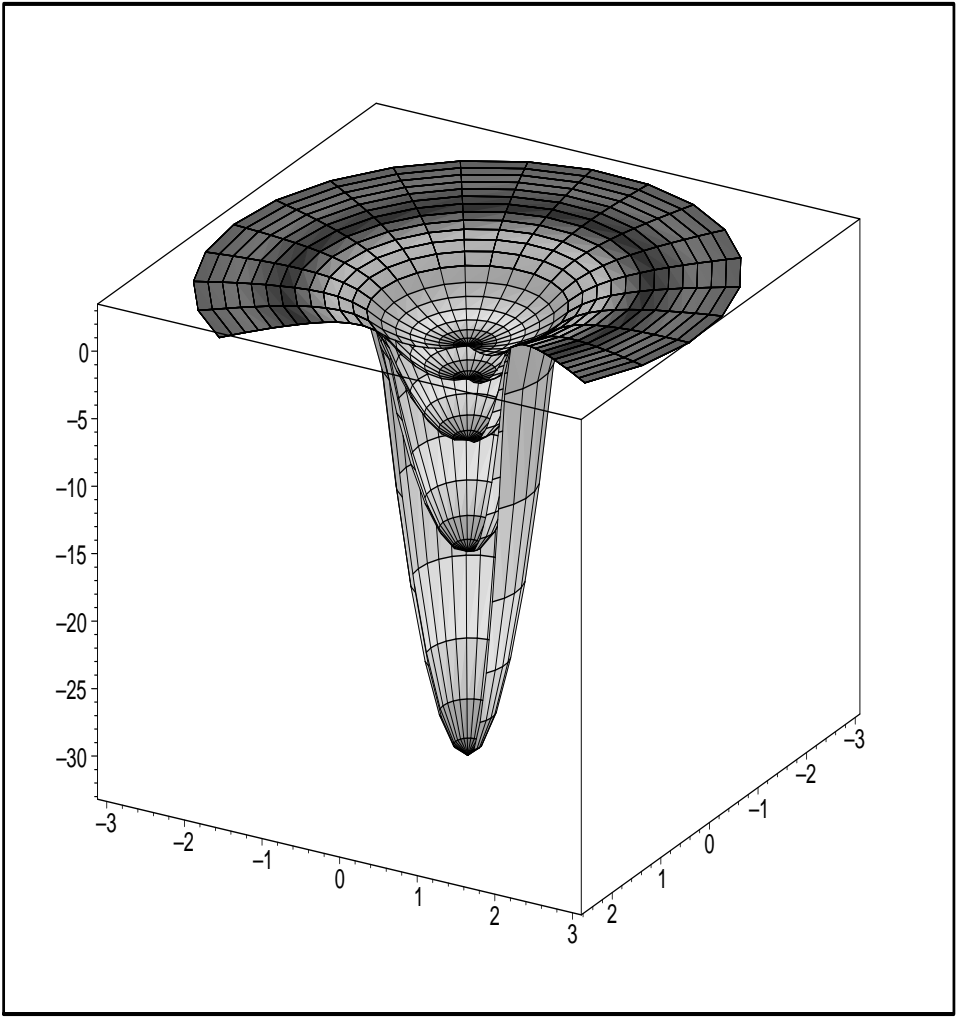}
  \caption{Dynamics of the isobar shape on the bottom edge of the condensation
area during formation of a cloud and development of the hydrodynamical
instability of a vortex (tornado). Different (descending)
shapes correspond to different equidistant times of vortex evolution.}
 \label{fig:tornado}
\end{center}
\end{figure}

In Fig. \ref{fig:tornado}. cylindrically-symmetric surfaces $ \tilde
z (r, t) = z (r, t)-z (\infty, 0) $ and their sections by mutually
perpendicular planes depending on $r $ for the isobar $P=P_{\mathrm
{evp}}, $ which corresponds to the bottom edge of a condensation
area of the moisture at $r\to \infty,$ are shown for 5 consecutive moments of time
with the identical intervals. They display the time
dynamics of a funnel evolution, which is filled with a fog (a
heterogeneous mixture "air --- water drops"), and develops during
the origination of a tornado on the bottom edge of a cloud. The
deepening of the funnel occurs till the contact of its minimum
that corresponds to the value $z_{\mathrm {min}} =-\tilde
g/\alpha^2 $ with the surface of the Earth, or till that moment of
time, when the maximum velocity of a vortex rotation reaches the
speed of sound.

Thus, the account for the gravity and vertical flows allows to describe
one of the main observable phenomenon --- a funnel formation on the
bottom edge of a cloud during the origination and development of
tornados.
%

\subsection{Discussion of viability of the obtained results for the
description of tornados and typhoons}

With the purpose of establishing of the viability of the obtained new
non-stationary solutions of the Navier-Stokes and continuity equations
for an incompressible viscous medium with a bulk sink and free inflow of
the substance for the description of atmospheric vortices (tornados
and typhoons), we will carry out numerical estimations of the
characteristic times of development of
hydrodynamical instabilities of vortical motion under conditions of
intensive condensation of moisture inside a cloud.

Let us consider a round cylindrical cloud of a radius $R_0\approx
1\div 10 $ km  slowly rotating in the twirled air flow with
equal by order initial values of azimuthal and radial velocities on
the border of a cloud $v_{r0} \approx v_{\varphi 0} \approx (1\div
10) $ m/s. This corresponds to the absolute values of initial angular
velocity $ \omega (0) $ and velocities of a converging radial flow
$\beta $ in the range $ (10^{-4} \div 10^{-2}) $ s$^{-1} $. For
humidity of atmospheric air about $100 \%,$ its density almost
twice exceeds the density of the dry air $ \rho_0=1.3\times 10^{-3} $
g/cm$^3$, so that the mentioned values of the parameter $ \beta $
are equivalent to capacities of a bulk sink $Q $ due to condensation
of a moisture of the order of $\sim 5\times (10^{-7} \div 10^{-5}) $
g/s$\cdot$cm$^3. $ Under these conditions, the characteristic time of
acceleration of a vortex rotation in the case of the exponential
instability equals $\tau=1/2\beta\approx (1\div 100)$ min.

The maximum velocity of the order of $c_s$ is reached in a time
interval of $t\approx (3\div 500)$ min. Typical observable times of
origination and development of a tornado indeed lie  in this time
interval (from several minutes to several hours).

For large-scale atmospheric vortices such as cyclones, hurricanes
and typhoons, which originate in cloud masses with sizes of about
$100$ km and more, characteristic times of the instability development
increase in two -- three orders of magnitude and can last for several
days, which also agrees with the observable times of hurricane and
typhoon existence. At the same time, the decreasing pressure on the
axis of a vortex caused by the cyclostrophic rotation regime explains
the characteristic "sucking"\ effect of a tornado.

Moreover, as marked above, in a two-phase system "air ---
water drops"\ the decrease of the pressure below the boiling point
of water at the given temperature should result in the termination of the process
of moisture condensation and lead to evaporation of the droplets. This
explains the formation of a clear tornado "core"\ or a typhoon
"eye"\ in the central part of a cloud system in the paraxial area of
a vortex. Inside this area the velocity of a vortical rotation of
the air slows down in time, since at the condition of $P <P_{\mathrm
{evp}}$  there is a bulk source of the gas
phase ($Q> 0$) instead of a bulk sink ($Q <0 $),
 and the parameter $ \beta $ changes its sign ($\beta
<0 $), that corresponds to the exponential deceleration of the air,
according to (\ref{2.8}). Such a condition of a dead calm is
observed in the center of the typhoon "eye".

It is also possible to estimate the moment of time $t_H$ when a tornado funnel contacts with
the surface of the Earth. In the case of the exponential instability, at
$z =-H $ (where $H $ is the height of the bottom edge of a cloud
above the ground), according to equations (\ref{3.33}) and
(\ref{2.8}), under the conditions $ \omega^2 (t) \gg \beta^2 $ and $
\tilde g/\alpha^2\gg H $ we find
\begin{equation}
t_H \approx \frac{1} {4\beta} \ln {\left[\frac{\tilde g H} {\omega^2 (0)
R_0^2} \right]}. \label{3.34}
\end{equation}
At $H=1$ km for the above mentioned values of $v_{\varphi 0} $ we
get the estimation of $t_H\approx (2\div 400) $ minutes.

In conclusion of this section it is necessary to emphasize, that the
growth of the kinetic energy and angular momentum of a vortex during the
development of the exponential or "explosive"\ instability does not
contradict to the conservation laws of energy and angular momentum,
because in an open system with a bulk sink and continuous inflow of
the substance, the transfer of the necessary amount of the kinetic energy
and angular momentum to the region $r\leqslant R_{\mathrm {max}} $
from the environment with a nonzero initial rotation velocity
$v_{\varphi 0} = \omega (0) R_0^2/r$ takes place. The value of the
maximum radius $R_{\mathrm {max}} $ in natural conditions at tornado
and typhoon origination on a free space of the land or sea is
actually limited only by a finite curvature of the Earth's surface.
This means that the maximum magnitudes of energy and momentum of
the atmospheric vortex cannot exceed the following values (per unit length):
\begin{equation}
\begin{array}{l}
\displaystyle E_{\mathrm {kin}} = \pi\rho\int_{R_0}^{R_{\mathrm
{max}}} r dr\left[v^2_{\varphi} (r) +v^2_{r} (r)
\right]\displaystyle =\pi\rho R_0^4\left[\omega^2 (0) + \beta^2
\right] \ln {\left( R_{\mathrm {max}}/R_0\right)}, \label{3.35}
\end{array}
\end{equation}
\begin{equation}
M_z=2\pi\rho\int_{R_0}^{R_{\mathrm {max}}} r^2 v_{\varphi} dr =
\pi\rho R_0^2R_{\mathrm {max}}^2\omega^2 (0). \label{3.36}
\end{equation}

For the above mentioned values of the parameters typical for
tornados and for $R_{\mathrm {max}} =10^3 $ km we get $E_{\mathrm
{max}} \approx (10^8\div 10^{12}) $ J/m. At this, the maximum
intensity liberated during the acceleration of a vortex
$\displaystyle\left(\frac{dE_{\mathrm {kin}}} {dt} \right)_{\mathrm
{max}} \sim 4\beta E_{\mathrm {max}}$ reaches roughly 1 GW per
meter of the vortex length, which explains the enormous destructive
force of a tornado.

Thus, the carried out numerical estimations show that the introduced in the paper a new
mechanism of hydrodynamical instability of a vortex in an
incompressible viscous medium under the action of convective and
Coriolis forces created by converging radial flows in open
thermodynamically non-equilibrium systems with a bulk sink and
unlimited inflow of a substance from the environment may be the
real cause of origination and development of powerful atmospheric
vortices --- tornados and typhoons --- during the intensive condensation
of water vapor from the humid air cooled below the dew-point during
formation of dense clouds.

\section{Conclusions}

In the present paper a new class of exact solutions of hydrodynamic
equations for an incompressible liquid (gas) at the presence of a bulk
sink and ascending flows of a substance was considered. For
existence of such solutions it is essential that one or several
components of a multicomponent fluid are excluded from the collective hydrodynamic
motion of the liquid at the expense of chemical or phase
transitions, but due to dynamical and chemical equilibrium in the
open system with the surrounding medium, a constant in time and
almost homogeneous in space chemical composition of the matter as well as
constant density $\rho=\mathrm{const}$ are maintained. It is shown that those profiles,
which nullify the terms in the Navier-Stokes equations that describe
viscous effects, exist and represent vortex structures with
"rigid-body"\ rotation of the core and converging radial flows.
In the case of constant bulk sink and inflow of the matter from the
outside, the azimuthal velocity of a "rigid-body"\ rotation
$v_\varphi$ increases exponentially in time. At simultaneous
infinite increase of the sink and inflow rates, $v_\varphi$
increases by a scenario of the "explosive"\ instability, where during
a finite time interval an infinite value of rotation velocity is reached.

On the basis of the developed theory of unstable hydrodynamical
vortices, the offered in \cite{5} mechanism of origination and
development of powerful atmospheric vortices --- tornados and
typhoons --- during the intensive condensation of water vapor from
the cooled below a dew point humid air during the formation of dense rain
clouds is considered. Within the framework of this mechanism it is
possible to explain the basic characteristics of tornados and
typhoons.  The characteristic times of
development of the instability of a vortical motion  agree in
order of magnitude with the corresponding times of origination and existence of
tornados (from several minutes to several hours) and typhoons
(several days).
The acceleration of a vortex rotation is limited by velocities
comparable with the speed of sound, when effects of the
compressibility of air, dissipative forces of a bulk viscosity and
large-scale turbulence start playing their roles.

With the account of gravity, the proposed model
describes the main phenomenon of a tornado --- the formation of a
lengthening funnel on the bottom edge of a cloud as a result of the
change of the shape of the surface of a constant pressure (isobar) that
bounds from the below the area of intensive condensation of
moisture. It is worth noting, that in an open system there are no
problems with the conservation laws of the energy and angular momentum,
because the radial flows transfer not only the necessary amount
of a substance, but also the necessary amount of the kinetic energy and
angular momentum from the slowly rotating environment.

The velocity jump on the border of a tornado core results in the
development of a strong turbulence, which can be described with the
help of the anomalous coefficient of turbulent viscosity, which in many
orders exceeds the usual viscosity of the air. Our estimations of
the turbulent viscosity coefficient are agreed with the values known
from the literature (see e.g. \cite{KKR2}).

\begin{acknowledgments}
\end{acknowledgments}

\end{document}